\documentclass[3p,review]{elsarticle}

\journal{Applied Mathematical Modelling}

\usepackage{amsmath,amssymb}
\usepackage{natbib}
\usepackage[shortlabels]{enumitem}

\usepackage{graphicx}
\usepackage{adjustbox}
\usepackage{multirow}
\usepackage{longtable}
\usepackage{pdflscape}
\usepackage{booktabs}

\usepackage[tableposition=above]{caption}
\begin{document}

\begin{frontmatter}

\title{Analysis of the One-dimensional Euler-Lagrange equation
of continuum mechanics with a Lagrangian of a special form}

\author[keldyshaddress,sutaddress]{E.I.~Kaptsov}
\ead{evgkaptsov@gmail.com}

\author[sutaddress]{S.V.~Meleshko\corref{corauthor}}
\cortext[corauthor]{Corresponding author}\ead{sergey@math.sut.ac.th}

\address[keldyshaddress]{Keldysh Institute of Applied Mathematics, \\
    Russian Academy of Science,
    Miusskaya Pl. 4, Moscow, 125047, Russia}

\address[sutaddress]{School of Mathematics, Institute of Science, \\
        Suranaree University of Technology, 30000, Thailand}

\begin{abstract}
Flows of one-dimensional continuum in Lagrangian coordinates are studied in the paper. Equations describing these flows are reduced to a single Euler-Lagrange
equation which contains two undefined functions. Particular choices of the undefined functions correspond to isentropic flows of an ideal gas, different forms of the hyperbolic shallow water equations.
Complete group classification of the equation with respect to these functions is  performed.

Using Noether's theorem, all conservation laws are obtained.
Their analogs in Eulerian coordinates are given.
\end{abstract}


\end{frontmatter}

\section{Introduction}

Modelling physical phenomena in continuum mechanics is considered
in two distinct ways. The typical approach uses Eulerian coordinates,
where flow quantities at each instant of time during motion are described
at fixed points. Alternatively, the Lagrangian description is used,
where the particles are identified by the positions which they occupy
at some initial time. Typically, Lagrangian coordinates are not applied
in the description of fluid motion.
In practice such description is often too detailed and complicated, but
it is always implied in formulating physical laws \cite{bk:Sedov[mss]}.
However, in some special contexts the Lagrangian description is indeed
useful in solving certain problems.

Many models in fluid mechanics presented in Eulerian coordinates can
be considered as a particular class of the model described by the
equations \cite{bk:GavrilyukShugrin[1996],bk:GavrilyukTeshukov2001}
\begin{equation}
\begin{array}{c}
\dot{\rho}+\rho\, \textrm{div}(u)=0,\ \ \rho\dot{u}+\nabla \partial=0,
\\[1.5ex]
\partial=\rho\frac{\delta W}{\delta\rho}-W=\rho \left(
    \frac{\partial W}{\partial\rho}-\frac{\partial}{\partial t}
        \left(\frac{\partial W}{\partial\dot{\rho}}\right)
        -\textrm{div}\left(\frac{\partial W}{\partial\dot{\rho}}u\right)
    \right) - W,
\end{array}
\label{eq:main}
\end{equation}
where $W\left(\rho,\dot{\rho}\right)$ is a given potential, `dot'
denotes the material time derivative: $\dot{f}=\frac{df}{dt}=f_{t}+u\nabla f$
and $\frac{\delta W}{\delta\rho}$ denotes the variational derivative
of $W$ with respect to $\rho$ at a fixed value of $u$. In particular,
the gas dynamics equations
and the shallow water equations have the form~(\ref{eq:main}).
Equations (\ref{eq:main}) can be derived as Euler-Lagrange equations in Lagrangian coordinates
\cite{bk:GavrilyukShugrin[1996],bk:GavrilyukTeshukov2001}.
The representation of equations (\ref{eq:main}) in the form of the Euler-Lagrange
equation allows one to use Noether's theorem for constructing conservation
laws \cite{bk:Ibragimov[1983]}.

In one-dimensional case Eulerian coordinates $(t,x)$ and mass Lagrangian
coordinates $(t,s)$ are related by the condition \cite{bk:Sedov[mss]}
\[
x=\varphi(t,s),
\]
where the function $\varphi(t,s)$ is a solution of the equation
\[
\varphi_{t}(t,s)=u(t,\varphi(t,s)).
\]
Here $u(t,x)$ is the velocity in Eulerian coordinates
which is related with the velocity $\tilde{u}(t,s)$
in the mass Lagrange coordinates by the relation
\cite{bk:Chernyi_gas}\footnote{Further we omit the sign $\tilde{ }$ as it is not ambiguous.}
\[
\tilde{u}(t,s)=u(t,\varphi(t,s)).
\]

In \cite{bk:SiriwatKaewmaneeMeleshko2016}
it was shown that one-dimensional
equations (\ref{eq:main}) are equivalent to an Euler-Lagrange
equation with the Lagrangian of the form
\[
{\cal L}=\frac{\varphi_{t}^{2}}{2} - \rho^{-1} W\left(\rho, \dot{\rho} \right),
\]
where $\rho=\varphi_{s}^{-1}$ and $\dot{\rho}=-\varphi_{s}^{-2}\varphi_{ts}$.

The one-dimensional hyperbolic shallow water equations with linear
bottom are derived from a particular case, where $W=W(\rho)$.
For a nonlinear bottom one has to assume that the function $W$
also depends on $x$.

The present paper is focused on Euler-Lagrange equations in Lagrangian coordinates, where
\[
    W = \tilde{g}(\rho) + \rho^{-1} h(x).
\]
The functions $\tilde{g}$ and $h$ are arbitrary functions of their arguments.
This case includes the hyperbolic shallow water equations in different forms,
and the isentropic gas dynamics equations.
Group properties of the latter equations in Eulerian coordinates were studied in
\cite{bk:Ovsiannikov1978,bk:ChirkunovDobrokhotovMedvedevMinenkov[2014],bk:SzatmariBihlo[2014],bk:AksenovDruzhkov[2016]}.
In
\cite{bk:HandbookLie_v1,bk:VinokurovNurgalieva[1985],bk:AndrKapPukhRod[1998],bk:AkhatovGazizovIbragimov[1991]}
group analysis was applied to the gas dynamics equations of a polytropic gas in mass Lagrangian coordinates.
Using Lagrangian and Noether's theorem, authors of \cite{bk:VinokurovNurgalieva[1985]} constructed
conservation laws for letter mentioned equations in isotropic case.
For nonlinear case the same approach was used in \cite{bk:WebbZank[2009]}.
Here we also should mention results of \cite{bk:SjobergMahomed2004},
where nonlocal conservation laws were obtained.

The objective of the present paper is to perform group classification of the Euler-Lagrange equation with respect
to the functions $g$, $h$, where the Lagrangian has the form
\begin{equation}
\label{lagr}
  L = \frac{\varphi_t^2}{2} + g\left(\varphi_s\right) + h(\varphi).
\end{equation}
Applying Noether's theorem, conservation laws are derived.

\section{The studied equation}

Considering the Lagrangian (\ref{lagr}),
one derives the Euler-Lagrange equation $\frac{\delta \mathcal{L}}{\delta \varphi} = 0$:
\begin{equation}  \label{el}
\varphi_{tt} + G \varphi_{ss} - H = 0.
\end{equation}
where $\frac{\delta}{\delta \varphi}$ is the variational derivative,
$G \equiv g^{\prime\prime}$,
and $H \equiv h^{\prime}$.
In order to preserve hyperbolicity of equation~(\ref{el}) it is assumed that $G < 0$.

In Eulerian coordinates, equation (\ref{el}) reduces to the system of equations
\begin{equation}
\label{ele}
\begin{array}{c}
    \rho_{t} + u \rho_{x}+\rho u_{x} = 0, \\
    u_{t} + u u_{x}-G\left(\frac{1}{\rho}\right) \rho^{-3} \rho_{x} - H(x) = 0.
\end{array}
\end{equation}

In particular, for equations studied in \cite{bk:SzatmariBihlo[2014]}
\[
    G\left( \frac{1}{\rho} \right) = -\gamma_1 \rho^2 \left( \rho + \gamma_2 \right),
    \quad
    H(x) = 0,
\]
and the second equation of (\ref{ele}) becomes
\[
    u_t + u u_x + \gamma_1 \left( 1 + \frac{\gamma_2}{\rho} \right) \rho_x = 0.
\]
Another particular case of equations (\ref{ele}) was analyzed in \cite{bk:AksenovDruzhkov[2016]},
where equations with a variable bottom $\chi(x)$ were considered.
For this case
\[
    G\left( \frac{1}{\rho} \right) = -\rho^3, \qquad
    H(x) = \chi^\prime(x),
\]
and the second equation of (\ref{ele}) becomes
\[
    u_t + u u_x + \rho_x - \chi^\prime = 0.
\]

\section{Group classification of equation (\ref{el})}

The first step in group classification consists of finding equivalence Lie group,
which can be used for simplifications of group classification.

\subsection{Equivalence Lie group}

Equation (\ref{el}) contains the arbitrary functions $G$ and $H$.
Equivalence transformations saving a structure of equations
are able to change the arbitrary elements.

A generator of an equivalence Lie group \cite{bk:Ovsiannikov1978}
is assumed to be in the form
\[
  X^e = \xi^t \partial_t + \xi^s \partial_s + \eta \partial_{\varphi}
    + \zeta^G \partial_{G}
    + \zeta^H \partial_{H},
\]
where $\xi^t$, $\xi^s$, $\eta$ depend on $(t,s,\varphi)$,
and $\zeta^G$, $\zeta^H$ depend on $(t,s,\varphi, G, H)$.

Applying the prolonged generator to the system
consisting of equation (\ref{el}) and the equations
\begin{eqnarray*}
    G_t = 0, \quad
    G_s = 0, \quad
    G_\varphi = 0, \quad
    G_{\varphi_t} = 0, \\
    H_t = 0, \quad
    H_s = 0, \quad
    H_{\varphi_s} = 0, \quad
    H_{\varphi_t} = 0,
\end{eqnarray*}
one obtains the following generators:
\begin{eqnarray}
\label{eq_trans}
    \begin{array}{l}
    X^e_1 = \partial_t,  \qquad
    X^e_2= \partial_s,  \qquad
    X^e_3 = \partial_\varphi,  \\
    X^e_4 = t \partial_t + s \partial_s - 2 H \partial_H - \varphi_s \partial_{\varphi_s}, \\
    X^e_5 = s \partial_s + 2 G \partial_G - \varphi_s \partial_{\varphi_s}, \\
    X^e_6 = \varphi \partial_\varphi + H \partial_H +  \varphi_s \partial_{\varphi_s}.
    \end{array}
\end{eqnarray}
These generators define the equivalence Lie group of equation (\ref{el}).

Notice that also there is the involution
\begin{equation}
\label{eq_trans_refl}
    \varphi \mapsto - \varphi, \quad
    H \mapsto - H,
\end{equation}
which is also an equivalence transformation.

In particular cases of arbitrary elements
the equivalence Lie group can be extended \cite{bk:Ovsiannikov1978}.
Extensions of the equivalence Lie group (\ref{eq_trans}) only
occur for particular choices of the function $H$.
These extensions are presented in Table~\ref{tab:eq_trans}.
Each column of the table lists the additional equivalence Lie group
generators for the corresponding particular functions $H$.

\begin{table}[htb]
\centering
\caption{Extensions of the equivalence Lie group}
\label{tab:eq_trans}
\begin{adjustbox}{max width=\textwidth}
\begin{tabular}{cccc}
\hline
$H=\alpha$
& $H=\alpha \varphi$, $\alpha \neq 0$
& $H=e^{\alpha \varphi}\!, \; \alpha \neq 0$
& $H=\beta \varphi^\alpha\!, \; \beta \neq 0$
\\
\hline
\hline
$\begin{array}{l} t \partial_\varphi, \\
s \partial_\varphi + \partial_{\varphi_s},\\
t^2 \partial_\varphi + 2 \partial_\alpha\\
\end{array}$
&
$\begin{array}{l}
\varphi \partial_\varphi + \varphi_s \partial_{\varphi_s},\\
\sinh(\sqrt{\alpha} t) \partial_\varphi \text{ if } \alpha > 0,\\
\cosh(\sqrt{\alpha} t) \partial_\varphi \text{ if } \alpha > 0,\\
\sin (\sqrt{|\alpha|} t) \partial_\varphi \text{ if } \alpha < 0, \\
\cos (\sqrt{|\alpha|} t) \partial_\varphi \text{ if } \alpha < 0\\
\end{array}$
& $\begin{array}{l}
t \partial_t + s \partial_s \\
\qquad - \frac{2}{\alpha} \partial_\varphi - \varphi_s \partial_{\varphi_s},\\
(\alpha \varphi + 1) \partial_\varphi - \alpha^2 \partial_\alpha  \\
\qquad + \alpha \varphi_s \partial_{\varphi_s}
\end{array}$
& $\begin{array}{l}
t \partial_t + s \partial_s
    - \frac{2 \varphi}{\alpha - 1} \partial_\varphi \\
    \qquad  + \frac{1+\alpha}{1-\alpha} \varphi_s \partial_{\varphi_s},\\
\beta \partial_\beta - \frac{\varphi}{\alpha - 1}\partial_\varphi
    - \frac{\varphi_s}{\alpha - 1} \partial_{\varphi_s}
\end{array}$
\\ \hline
\end{tabular}%
\end{adjustbox}
\end{table}

In the group classification the following equivalence transformations are used:
the transformation corresponding to the generator $X^e_3$
\[
\varphi \mapsto \varphi + \varepsilon_1;
\]
the transformation corresponding to the generator $X^e_4$
\[
t \mapsto t e^{\varepsilon_2}, \quad
s \mapsto s e^{\varepsilon_2}, \quad
H \mapsto H e^{-2 \varepsilon_2};
\]
the transformation corresponding to generator
$t^2 \partial_\varphi + 2 \partial_\alpha$
\[
\varphi \mapsto \varphi + \varepsilon_3 t^2, \quad
\alpha \mapsto \alpha + 2 \varepsilon_3;
\]
and the transformation corresponding to generator
$s \partial_\varphi + \partial_{\varphi_s}$,
which allows one to shift $\varphi_s$,
\[
s \mapsto s \varphi + \varepsilon_4 s, \qquad
\varphi_s \mapsto \varphi_s + \varepsilon_4.
\]

\subsection{Admitted Lie group}

A generator admitted by equation (\ref{el}) is considered in the form
\[
  X = \xi^t \partial_t + \xi^s \partial_s + \eta \partial_\varphi,
\]
where its coefficients depend on $(t, s, \varphi)$.
\[
    \xi^s = C_1 s + C_2, \qquad
    \eta = \left(\frac{1}{2} \xi^t_t + C_3\right) \varphi + \tau_1 s + \tau_2,
\]
where the constants $C_1$, $C_2$, $C_3$
and the functions $\tau_1(t)$ , $\tau_2(t)$ and $\xi^t = \xi^t(t)$ satisfy the classifying equations
\begin{equation}\label{rsys4}
\def\arraystretch{1.25}
  \begin{array}{l}
    \left( \left( \frac{1}{2} \xi^t_t + C_3 - C_1 \right) \varphi_s + \tau_1 \right) G^\prime + 2 (\xi^t_t - C_1) G = 0, \\
    \left( \left( \frac{1}{2} \xi^t_t + C_3 \right) \varphi + \tau_2 \right) H^\prime
        - \left( C_3 - \frac{3}{2} \xi^t_t \right) H
        - \frac{1}{2} \xi^t_{ttt} \varphi - \tau_2^{\prime\prime} = 0,
    \\
    \tau_1 H^\prime - \tau_1^{\prime\prime} = 0.
  \end{array}
\end{equation}
Notice that here and further on $C_i$ are constant.

The kernel of admitted Lie algebras is derived by
splitting equations (\ref{rsys4}) with respect to $G$, $G^\prime$, $H$ and $H^\prime$.
Its basis consists of the generators
\begin{equation}\label{kern}
  X_1 = \partial_t, \qquad
  X_2 = \partial_s.
\end{equation}
Extensions of the kernel of admitted Lie algebras occur for particular cases of the functions $G$ and $H$. Further these particular cases are derived.

Differentiating the third equation of system (\ref{rsys4}) by $\varphi$, one obtains
\[
    \tau_1 H^{\prime\prime} = 0
\]
Hence, the analysis of system (\ref{rsys4}) is split into the cases: a linear function $H$, i.e. $H^{\prime\prime}=0$,
and a nonlinear function $H$, i.e. $\tau_1 = 0$.

\subsection{The function $H$ is linear}

For linear function $H$, i.e.,
\[
H = \alpha \varphi + \beta,
\]
where $\alpha$ and $\beta$ are constant, one can assume that $\beta = 0$.
Indeed, if $\alpha = 0$, then, using the equivalence transformations corresponding to the generator
$t^2 \partial_\varphi + 2 \partial_\beta$, one can assume that $\beta = 0$.
If $\alpha \neq 0$ then one can reduce $\beta$ by the equivalence
transformations corresponding to the generator $X^e_3$.

\subsubsection{Case of $H = 0$}

Substituting $H = 0$ in (\ref{rsys4}), one obtains
\[
    \tau_1^{\prime\prime} = 0, \qquad
    \tau_2^{\prime\prime} = 0, \qquad
    \xi^t_{ttt} = 0.
\]
It follows that
\[
    \xi^t = K_1 t^2 + K_2 t + K_3, \qquad
    \tau_1 = A_1 t + A_2, \qquad
    \tau_2 = B_1 t + B_2,
\]
where $A_i$, $B_i$, $K_i$ are constant.
Substituting the latter relations into (\ref{rsys4}) and splitting by $t$, one derives that
\begin{equation}
    \label{H0_sys1}
    \def\arraystretch{1.25}
    \begin{array}{c}
        (K_1 \varphi_s + A_1) G^\prime + 4 K_1 G = 0, \\
        \left( \left(
            \frac{1}{2} K_2 + C_3 - C_1 \right) \varphi_s + A_2
        \right) G^\prime + 2 (K_2 - C_1) G = 0.
    \end{array}
\end{equation}
For an arbitrary function $G$ one obtains that
\[
    K_1 = 0, \quad
    A_1 = 0, \quad
    A_2 = 0, \quad
    K_2 = C_1 = 2 C_3,
\]
and the extension of the kernel (\ref{kern}) is defined by the generators
\begin{equation} \label{H0_kern}
  \partial_\varphi, \quad
    t \partial_\varphi, \quad
    t \partial_t + s \partial_s + \varphi \partial_\varphi.
\end{equation}

From system (\ref{H0_sys1}) one can conclude that there exist constants $a$, $b$ and $c$ such that
\begin{equation}\label{H0_sys1a}
    (a \varphi_s + b) G^\prime + c G = 0.
\end{equation}
By virtue of the equivalence transformations, for finding extensions of (\ref{H0_kern})
one needs to study two representations of the functions $G$.

The first form of $G$ is
\[
    G = -e^{\mu \varphi_s},
\]
where $\mu \neq 0$ is constant.
Substituting $G$ into (\ref{H0_sys1}), one obtains that
\[
    K_1 = 0, \quad
    A_1 = 0, \quad
    K2 = 2 (C_1 - C_3), \quad
    2 (C_1 - 2 C_3) + \mu A_2 = 0.
\]
The extension of (\ref{H0_kern}) is defined by the generator
\[
    t \partial_t - \frac{2 s}{\mu} \partial_\varphi.
\]

The second form of the function $G$ is defined by the generators
\[
    G = -\varphi_s^\lambda,
\]
where $\lambda$ is constant such that $\lambda (1 - \lambda) \neq 0$.
Substituting $G$ into (\ref{H0_sys1}), one obtains that
\[
    A_1 = 0, \quad
    A_2 = 0, \quad
    K_1 (\lambda + 4) = 0, \quad
    (\lambda + 4) K_2 - 2 C_1 (\lambda + 2) + 2 \lambda C_3 = 0.
\]

If $\lambda = -4$, then $C_1 = 2 C_3$, and the extension of (\ref{H0_kern}) is
\[
    2 t \partial_t + \varphi \partial_\varphi \quad
    t^2 \partial_t + t \varphi \partial_\varphi.
\]

If $\lambda = -2$, then $K_1 = 0$ and $K_2 = 2 C_3$.
The extension of (\ref{kern}) is
\[
    s \partial_s, \quad
    t \partial_t + \varphi \partial_\varphi.
\]

If $(\lambda + 4)(\lambda + 2) \neq 0$, then
\[
    K_1 = 0, \quad
    K_2 = \frac{2 C_1 (\lambda + 2) - 2 \lambda C_3}{\lambda + 4},
\]
and the extension of (\ref{H0_kern}) is
\[
    t \partial_t - \frac{2}{\lambda} \varphi \partial_\varphi.
\]

\subsubsection{Case of $H = \alpha \varphi$, $\alpha \neq 0$}

It follows from system (\ref{rsys4}) that
\begin{equation}
    \label{rsys_linH}
    \begin{array}{l}
        \left(\left(
                \frac{1}{2} \xi^t_{t} + C_3 - C_1
            \right) \varphi_s
            + \tau_1 \right) G^\prime + 2 (\xi^t_{t} - C_1) G = 0,
        \\
        \xi^t_{ttt} = 4 \alpha \xi^t_t, \quad
        \tau_1^{\prime\prime} = \alpha \tau_1   \quad
        \tau_2^{\prime\prime} = \alpha \tau_2.
    \end{array}
\end{equation}

If $G$ is an arbitrary function, then
\[
\xi^t_t = C_1 = C_3 = 0, \quad \tau_1 = 0,
\]
The extension of (\ref{kern}) is defined by the generators
\begin{equation}
    \label{trig_ext_1}
    \def\arraystretch{1.25}
    \begin{array}{c}
          \sin \sqrt{|\alpha|} t \partial_\varphi, \quad
          \cos \sqrt{|\alpha|} t \partial_\varphi,
          \quad \alpha < 0,
          \\
          \cosh \sqrt{\alpha} t \partial_\varphi, \quad
          \sinh \sqrt{\alpha} t \partial_\varphi, \quad \alpha > 0.
    \end{array}
\end{equation}
As in the previous case, one has to study two forms of the function $G$.
The first form is
\[
    G = -e^{\mu \varphi_s},
\]
where $\mu \neq 0$ is constant.
Substituting $G$ into (\ref{rsys4}), one gets
\[
    C_1 = C_3 = 0, \quad
    \xi^t_t = 0, \quad
    \tau_1 = 0, \quad
    \tau_2^{\prime\prime} - \alpha \tau_2 = 0.
\]
In this case there is no an extension of (\ref{trig_ext_1}).

Another form of $G$ is
\[
    G = -(\varphi_s + c)^\lambda,
\]
where $\lambda$ and $c$ are constant, such that $\lambda (1 - \lambda) \neq 0$.
Substituting $G$ into the first equation of (\ref{rsys4}), one obtains
\[
    (\lambda + 4) \xi^t_t - 2 (\lambda + 2) C_1 + 2 \lambda C_3 = 0, \quad
    \tau_1 = \frac{2 c (\xi^t_t - C_1)}{\lambda}, \quad
    c (3 \xi^t_t + C_1) = 0.
\]

If $c \neq 0$ then
\[
    \xi^t_t = 0, \quad
    \tau_1 = 0, \quad
    C_1 = 0, \quad
    C_3 = 0,
\]
and there is no an extension of (\ref{trig_ext_1}) in this case.

If $c = 0$, then $\tau_1 = 0$.
Integrating the last three equations of system (\ref{rsys_linH}), one obtains
\begin{equation}\label{harm_rels}
\begin{array}{l}
\xi^t = A_3 +
\begin{cases}
          A_1 \sin 2\sqrt{|\alpha|} t + A_2 \cos 2\sqrt{|\alpha|} t, \quad \alpha < 0, \\
          A_1 \sinh 2\sqrt{\alpha} t + A_2 \cosh 2\sqrt{\alpha} t, \quad \alpha > 0;
        \end{cases} \\
    \tau_2 = \begin{cases}
          B_1 \sin \sqrt{|\alpha|} t + B_2 \cos \sqrt{|\alpha|} t, \quad \alpha < 0, \\
          B_1 \sinh \sqrt{\alpha} t + B_2 \cosh \sqrt{\alpha} t, \quad \alpha > 0,
        \end{cases}
\end{array}
\end{equation}
where $A_i$, $B_i$ are constant.
Substituting (\ref{harm_rels}) into the
first equation of system (\ref{rsys_linH}), one derives that
\[
B_1 = 0, \quad
B_2 = 0, \quad
\sqrt{|\alpha|} (\lambda + 4) A_1 + C_1 (2 + \lambda) - \lambda C_3 = 0,
\]
and
\[
    \begin{array}{c}
    (\lambda + 4)\sqrt{|\alpha|} (A_2 \sin 2 \sqrt{|\alpha|}t - 2 A_1 \cos^2 \sqrt{|\alpha|}t) = 0,
    \quad
    \alpha < 0;
    \\
    (\lambda + 4)\sqrt{\alpha} (A_2 \sinh 2 \sqrt{\alpha}t - 2 A_1 \cosh^2 \sqrt{\alpha}t) = 0,
    \quad
    \alpha > 0.
    \end{array}
\]
Then, for any $\lambda \neq 0$ the following extension of (\ref{trig_ext_1}) occurs
\[
    s \partial_s + \frac{2 + \lambda}{\lambda} \varphi \partial_\varphi.
\]
If $\lambda = -4$, then there are additional to (\ref{trig_ext_1}) generators
\[
    \begin{array}{c}
    \sin 2 \sqrt{|\alpha|} t \partial_t
        + {\sqrt{|\alpha|} \varphi} \cos 2 \sqrt{|\alpha|} t \partial_\varphi,
        \quad \alpha < 0;\\
    \sinh 2 \sqrt{\alpha} t \partial_t
        + \sqrt{\alpha} \varphi \cosh 2 \sqrt{\alpha} t \partial_\varphi,
        \quad \alpha > 0,
    \end{array}
\]
and
\[
    \begin{array}{c}
    \cos 2 \sqrt{|\alpha|} t \partial_t
        - \sqrt{|\alpha|} \varphi \sin 2 \sqrt{|\alpha|} t \partial_\varphi,
        \quad \alpha < 0;\\
    \cosh 2 \sqrt{\alpha} t \partial_t
        + \sqrt{\alpha} \varphi \sinh 2 \sqrt{\alpha} t \partial_\varphi,
        \quad \alpha > 0.
    \end{array}
\]

\subsection{The function $H$ is nonlinear}

As mentioned for the nonlinear function $H$ one has that $\tau_1 = 0$.
System (\ref{rsys4}) becomes
\begin{equation}\label{rsys4nlin}
\def\arraystretch{1.25}
  \begin{array}{l}
    \left( \frac{1}{2} \xi^t_t + C_3 - C_1 \right) \varphi_s \frac{G^\prime}{G} + 2 (\xi^t_t - C_1) = 0, \\
    \left( \left( \frac{1}{2} \xi^t_t + C_3 \right) \varphi + \tau_2 \right) H^\prime
        - \left( C_3 - \frac{3}{2} \xi^t_t \right) H
        - \frac{1}{2} \xi^t_{ttt} \varphi - \tau_2^{\prime\prime} = 0.
  \end{array}
\end{equation}
Recall that
\[
    \xi^t = \xi^t(t), \qquad
    \xi^s = C_1 s + C_2, \qquad
    \eta = \left(\frac{1}{2} \xi^t_t + C_3\right) \varphi + \tau_2(t).
\]
Differentiating the first equation of system (\ref{rsys4nlin}) with respect to $\varphi_s$, one obtains
\[
    \left(\frac{1}{2} \xi^t_t + C_3 - C_1 \right) \left( \varphi_s \frac{G^\prime}{G} \right)^\prime = 0.
\]

\paragraph{Case of $\left( \varphi_s \frac{G^\prime}{G} \right)^\prime \neq 0$}
The latter equation and the first equation of (\ref{rsys4nlin}) gives that
\[
    \xi^t_t = C_1, \quad
    C_3 = C_1 / 2.
\]
The second equation of (\ref{rsys4nlin}) becomes
\[
  \left( C_1 \varphi + \tau_2 \right) H^\prime
    + C_1 H - \tau_2^{\prime\prime}= 0.
\]
Differentiating the latter equation by $\varphi$ and then by $t$,
one gets that $\tau_2^\prime H^{\prime\prime}=0$, which gives that $\tau_2$ is constant, and
\begin{equation}
\label{rsys_nlin}
    \left( C_1 \varphi + \tau_2 \right) H^\prime + C_1 H = 0,
\end{equation}

Notice, that, if $C_1 = 0$, then for the nonlinear $H$, one has that $\tau_2 = 0$,
which does not lead to an extension of the kernel (\ref{kern}).
Hence, one needs to consider the case $C_1 \neq 0$.
By virtue of the equivalence transformations one can assume that the function $H$ has the form
\[
H = \beta \varphi^{-1},
\]
where $\beta \neq 0$ is constant.
Substituting the latter relation into (\ref{rsys_nlin}), one derives that $\tau_2 = 0$,
and $G$ is an arbitrary nonlinear function.
Thus, for $\left( \varphi_s \frac{G^\prime}{G} \right)^\prime \neq 0$
the extension of (\ref{kern}) is given by the generator
\[
    t \partial_t + s \partial_s + \varphi \partial_\varphi.
\]

\paragraph{Case of $\left( \varphi_s \frac{G^\prime}{G} \right)^\prime = 0$}
Here
\[
G = -(\varphi_s + c)^\lambda,
\]
where $c$ and $\lambda \neq 0$ are constant.
By virtue of the nonlinearity of $G$, one has that $(\lambda - 1) \lambda \neq 0$.

Substituting the latter relation into the first equation of system (\ref{rsys4nlin}), one derives
the following system of equations
\begin{equation}\label{lambda_rels}
    \begin{array}{c}
        c (\xi^t_t - C_1) = 0, \\
        (\lambda + 4) \xi^t_t - 2 C_1 (\lambda + 2) + 2 \lambda C_3 = 0.
    \end{array}
\end{equation}

If $c \neq 0$, then $\xi^t_t = C_1$, $C_3 = C_1 / 2$,
and the equation for $H$ becomes (\ref{rsys_nlin}).
This case was studied above.

Hence, one needs to consider $c = 0$, i.e. the function $G = -\varphi^\lambda$.

Differentiating the second equation of system (\ref{rsys4nlin})
twice by $\varphi$, and then by $t$, one gets
\begin{equation}
    \label{dec6_star_1}
    5 \xi^t_{tt} H^{\prime\prime}
        + (\xi^t_{tt} \varphi + 2 \tau_2^\prime) H^{\prime\prime\prime} = 0.
\end{equation}
First we analyze the case $H^{\prime\prime\prime} \neq 0$.
Differentiating (\ref{dec6_star_1}) by $\varphi$, one obtains
\begin{equation}
    \label{dec6_star_2}
  \xi^t_{tt} \left (
        5 \left( \frac{H^{\prime\prime}}{H^{\prime\prime\prime}} \right)^\prime
        + 1
    \right) = 0.
\end{equation}
Thus, one ends up with the following cases.

\subparagraph{Case $\left( \frac{H^{\prime\prime}}{H^{\prime\prime\prime}} \right)^\prime = -1/5$.}

Integrating one finds that
\[
H = \alpha (\varphi + \delta)^{-3} + \beta \varphi + \varepsilon,
\]
where $\alpha \neq 0$, $\beta$ and $\varepsilon$, $\delta$ are constant.
By virtue of the equivalence transformation corresponding to the generator $X^e_3$
one can assume $\delta$ = 0.

Substituting $H$ into system (\ref{rsys4nlin}), one gets
\begin{equation}
    \label{dec6_star_3}
    \def\arraystretch{1.25}
    \begin{array}{c}
        (\lambda + 4) \xi^t_{tt} + 2 (C_3 - C_1) \lambda - 4 C_1 = 0, \\
        (\xi^t_{ttt} - 4 \beta \xi^t_t) \varphi
         + 2 \alpha ( 4 C_3 \varphi^{-3} + 3 \tau_2 \varphi^{-4} )
         + \varepsilon (2 C_3 - 3 \xi^t_t) + 2 (\tau_2^{\prime\prime} - \beta \tau_2) = 0.
    \end{array}
\end{equation}
Splitting the latter equation with respect to $\varphi$,
system of equations (\ref{lambda_rels}), (\ref{dec6_star_3}) reduces to the equations
\begin{equation}\label{case_1_lam4}
    \def\arraystretch{1.25}
    \begin{array}{c}
        C_3 = 0, \quad
        \tau_2 = 0, \\
        (\lambda + 4) \xi^t_t - 2 (\lambda + 2) C_1  = 0, \\
        \xi^t_{ttt} - 4 \beta \xi^t_t = 0, \qquad
        \varepsilon \xi^t_t = 0.
    \end{array}
\end{equation}
Notice that if $\varepsilon \neq 0$, then extension of (\ref{kern}) only occurs for $\lambda = -2$,
and it is defined by the generator $s \partial_s$.
Hence, it is assumed that $\varepsilon = 0$.

\begin{enumerate}[(i)]
  \item If $\lambda = -4$, then $C_1 = 0$, and system (\ref{case_1_lam4}) gives that $\xi^t$ must satisfy the following equation
  \[
        \xi^t_{ttt} - 4 \beta \xi^t_t = 0.
  \]
  If $\beta = 0$, then $\xi^t_{ttt} = 0$ and one derives that the extension of (\ref{kern}) is
  \[
    \partial_\varphi, \quad
    t \partial_\varphi, \quad
    2 t \partial_t + \varphi \partial_\varphi \quad
    t^2 \partial_t + 2 t \varphi \partial_\varphi.
  \]
  If $\beta \neq 0$, then the extensions of (\ref{kern}) are
  \[
      \begin{array}{c}
        \sin 2 \sqrt{|\beta|} t \partial_t
            + {\sqrt{|\beta|} \varphi} \cos 2 \sqrt{|\beta|} t \partial_\varphi,
            \; \beta < 0,\\
        \sinh 2 \sqrt{\beta} t \partial_t
            + \sqrt{\beta} \varphi \cosh 2 \sqrt{\beta} t \partial_\varphi,
            \; \beta > 0,
      \end{array}
  \]
  and
  \[
        \begin{array}{c}
        \cos 2 \sqrt{|\beta|} t \partial_t
            - \sqrt{|\beta|} \varphi \sin 2 \sqrt{|\beta|} t \partial_\varphi,
            \; \beta < 0,\\
        \cosh 2 \sqrt{\beta} t \partial_t
            + \sqrt{\beta} \varphi \sinh 2 \sqrt{\beta} t \partial_\varphi,
            \; \beta > 0.
        \end{array}
  \]
  Notice, that for $H = \alpha \varphi^{-3} + \beta \varphi$ and $G = -\varphi^{-4}$
  the extensions coincide with the case where $\alpha = 0$.

  \item
  If $\lambda = -2$, then the extension of (\ref{kern}) is defined by the generator $s \partial_s$.

  \medskip

  \item If $(\lambda + 2)(\lambda + 4) \neq 0$,
  then, form (\ref{case_1_lam4}), one gets
  \[
    \xi^t_t = 2 C_1 \frac{\lambda + 2}{\lambda + 4}, \quad
    \beta C_1 (\lambda + 2) = 0.
  \]
  The extension of (\ref{kern}) only occurs for $\beta = 0$:
  \[
    2 t \partial_t + \frac{\lambda + 4}{\lambda + 2} s \partial_s + \varphi \partial_\varphi.
  \]
\end{enumerate}

\subparagraph{Case of $\left( \frac{H^{\prime\prime}}{H^{\prime\prime\prime}} \right)^\prime \neq -1/5$.}

Recall that in this case $\xi^t_{tt} = 0$, i.e.,
\[
    \xi^t = A_1 t + A_2,
\]
where $A_i$ are constant.
System (\ref{rsys4nlin}) becomes
\begin{equation}\label{nlin_case2_sys}
    \begin{array}{c}
    (\lambda + 4) A_1 - 2 (\lambda + 2) C_1 + \lambda C_3 = 0, \\
    ((A_1 + 2 C_3) \varphi + 2 \tau_2) H^{\prime} + (3 A_1 - 2 C_3) H - 2 \tau_2^{\prime\prime} = 0.
    \end{array}
    \end{equation}
Differentiating the second equation of (\ref{nlin_case2_sys})
with respect to $\varphi$ and $t$, one obtains that $\tau_2$ is constant,
and analysis of the second equation of (\ref{nlin_case2_sys}) leads to the study of two types of the function:
$H = e^{\alpha \varphi}$, where $\alpha \neq 0$,
and $H = \beta \varphi^\alpha$, where $\alpha (1 - \alpha) \neq 0$.

\begin{enumerate}[(i)]

  \item If $H = e^{\alpha \varphi}$, then system (\ref{nlin_case2_sys}) reduces to the equations
    \[
        (\lambda + 2) C_1 + 4 C_3 = 0, \quad
        A_1 = -2 C_3, \quad
        \alpha \tau_2 = 4 C_3.
    \]

    If $\lambda \neq -2$, then $\tau_2 = 4 C_3 / \alpha$ and $C_3 = -C_1 (\lambda + 2) / 4$.
    Hence, the extension of (\ref{kern}) is
    \[
        t \partial_t + \frac{2}{2 + \lambda} s \partial_s - \frac{2}{\alpha} \partial_\varphi.
    \]

    If $\lambda = -2$, then $C_3 = 0$ and $\tau_2 = 0$,
    and the only extension of the kernel (\ref{kern}) is given by the generator\,$s \partial_s$.

  \medskip

  \item In case of $H = \beta \varphi^\alpha$,
  one notices that from the condition
  $\left( \frac{H^{\prime\prime}}{H^{\prime\prime\prime}} \right)^\prime \neq -1/5$ implies that $\alpha \neq -3$.

  The second equation of system (\ref{nlin_case2_sys}) leads to the equations
  \[
        \tau_2 = 0, \quad
        A_1 = 2 \frac{1 - \alpha}{3 + \alpha} C_3.
  \]

    If $\lambda \neq -2$, then
    \[
        C_1 = \frac{4 C_3 (\lambda - \alpha + 1)}{(\lambda + 2)(\alpha + 3)},
    \]
    and the extension of the kernel (\ref{kern}) is defined by the generator
    \begin{equation}\label{lambd_gen}
      t \partial_t
            + \frac{2 (\alpha  - \lambda - 1)}{(2 + \lambda)(\alpha - 1)} s \partial_s
            + \frac{2 \varphi}{1 - \alpha} \partial_\varphi.
    \end{equation}

    \medskip

    If $\lambda = -2$, then the first equation of system (\ref{nlin_case2_sys}) reduces to
    \[
        C_3 (\alpha + 1) = 0.
    \]
    If $\alpha \neq -1$, then $C_3 = 0$, and the extension of the kernel (\ref{kern}) is
    defined by the generator $s \partial_s$.
    If $\alpha = -1$,  then the extension of (\ref{kern})  is
    \[
        s \partial_s, \quad
        t \partial_t + \varphi \partial_\varphi.
    \]

\end{enumerate}

\subparagraph{Case of $H^{\prime\prime\prime} = 0$.}

  In this case it follows that
  \[
  H = \alpha \varphi^2 + \beta \varphi + \delta,
  \]
  where $\alpha \neq 0$, $\beta$, $\delta$ are constant,
  and the constant $\delta$ can be vanished by equivalence transformations.

  Splitting the second equation of (\ref{rsys4nlin}), and using the second equation of (\ref{lambda_rels}),
  one obtains
  \[
    \xi^t_t = 2 \frac{\lambda + 2}{1 - \lambda} C_1, \quad
    C_3 = -5 \frac{\lambda + 2}{1 - \lambda} C_1, \quad
    \tau_2 = 0.
  \]
  In this case the extension of the kernel (\ref{kern}) is defined by the generator
  \[
        (\lambda + 2)(t \partial_t - 2 \varphi \partial_\varphi) + 2 (1 - \lambda) s \partial_s.
  \]
  Notice that this is just a particular case of (\ref{lambd_gen}).

\bigskip


Results of the presented above classifications are given in
Table~\ref{tab:syms_arb} and Table~\ref{tab:syms}.
Table~\ref{tab:syms_arb} contains extensions of the kernel of admitted Lie algebras,
where one of the functions, either $G$ or $H$, is arbitrary.
The columns of the table correspond to different functions $H$,
and the rows correspond to different functions $G$.
On the intersection of the columns and the rows the extensions of~(\ref{kern}) are presented.
Generators provided in Table~\ref{tab:syms} extend the contents of Table~\ref{tab:syms_arb}.
Constraints on the values of constants are included in the table.


\begin{table}[htb]
\centering
\caption {Extensions of (\ref{kern}) (arbitrary $G$ or $H$).}
\label{tab:syms_arb}
\begin{adjustbox}{max width=\textwidth}
\begin{tabular}{c|cccc}
\hline
& $H=H(\varphi)$
& $H= 0$
& $H=\alpha \varphi$, $\alpha \neq 0$
& $H=\beta \varphi^{-1}$, $\beta \neq 0$
\\
\hline
\hline
$G=G(\varphi_s)$
&
$\begin{array}{l}
X_1 = \partial_t, \\
X_2 = \partial_s \\
\end{array}$
&
$\begin{array}{l}
X_3 = \partial_\varphi, \\
X_4 = t \partial_\varphi, \\
X_5 = t \partial_t + s \partial_s + \varphi \partial_\varphi\\
\end{array}$
&
$\begin{array}{l}
\\
X_3 = \begin{cases}
    \sin \sqrt{|\alpha|} t \partial_\varphi, \quad
        \alpha < 0,\\
    \sinh \sqrt{\alpha} t \partial_\varphi, \quad
        \alpha > 0,
    \end{cases}
    \\
X_4 = \begin{cases}
    \cos{\sqrt{|\alpha|} t} \partial_\varphi, \quad
        \alpha > 0,\\
    \cosh{\sqrt{\alpha} t} \partial_\varphi, \quad
        \alpha > 0.
    \end{cases}
\end{array}$
&
$\begin{array}{l}
X_3 = t \partial_t + s \partial_s + \varphi \partial_\varphi.
\end{array}$
\\
\hline
$G=-\varphi_s^{-2}$
&
$X_3 = s \partial_s$
& \multicolumn{3}{c}{--}
\\
\hline
\end{tabular}
\end{adjustbox}
\end{table}

\begin{table}[htb]
\centering
\caption {Extensions of (\ref{kern}).}
\label{tab:syms}
\begin{adjustbox}{max width=\textwidth}
\begin{tabular}{c|ccccc}
\hline
& $H=0$
& $\begin{array}{c}H=\beta \varphi^{-3} + \alpha \varphi, \\ \alpha \neq 0\end{array}$
& $\begin{array}{c} H=e^{\alpha \varphi}, \\ \alpha \neq 0 \end{array}$
& $\begin{array}{c} H=\beta \varphi^{\alpha}, \\ \alpha \neq 0, 1, \; \beta \neq 0 \end{array}$ \\
\hline
\hline
$\begin{array}{l}
G=-e^{\mu \varphi_s}, \\
\mu \neq 0
\end{array}$
&
$
X_6 = t \partial_t - \frac{2}{\mu} s \partial_\varphi
$
&
--
& --
& --
\\
\hline
$\begin{array}{c}
G=-\varphi_s^{\lambda} \\
\lambda \neq 0, 1
\end{array}$
&
$\begin{array}{l}
X_6 = t \partial_t
    -\frac{2}{\lambda} \varphi \partial_\varphi \\
\text{if } \lambda = -4: \\
X_7 = t^2 \partial_t + t \varphi \partial_\varphi
\end{array}$
&
$\begin{array}{l}
    \text{if } \beta = 0: \\
X_5 = s \partial_s + \frac{1}{\lambda} (2 + \lambda) \varphi \partial_\varphi. \\
    \text{if } \lambda = -4: \\
X_6 = \begin{cases}
    \sin 2 \sqrt{|\alpha|} t \partial_t
        + {\sqrt{|\alpha|} \varphi} \cos 2 \sqrt{|\alpha|} t \partial_\varphi,
        \; \alpha < 0,\\
    \sinh 2 \sqrt{\alpha} t \partial_t
        + \sqrt{\alpha} \varphi \cosh 2 \sqrt{\alpha} t \partial_\varphi,
        \; \alpha > 0,
    \end{cases}
    \\
X_7 = \begin{cases}
    \cos 2 \sqrt{|\alpha|} t \partial_t
        - \sqrt{|\alpha|} \varphi \sin 2 \sqrt{|\alpha|} t \partial_\varphi,
        \; \alpha < 0,\\
    \cosh 2 \sqrt{\alpha} t \partial_t
        + \sqrt{\alpha} \varphi \sinh 2 \sqrt{\alpha} t \partial_\varphi,
        \; \alpha > 0.
    \end{cases}
\end{array}$
&
$\begin{array}{l}
    \text{if } \lambda \neq -2: \\
    X_3 = t \partial_t \\
    \quad + \frac{2}{2 + \lambda} s \partial_s \\
    \qquad - \frac{2}{\alpha} \partial_\varphi.
 \end{array}
$
&
$\begin{array}{l}
\text{if } \lambda \neq -2: \\
X_3 = t \partial t \\
    \quad + \frac{2 (\alpha - \lambda - 1)}{(2+\lambda)(\alpha - 1)} s \partial_s \\
    \quad + \frac{2 \varphi}{1 - \alpha} \partial_\varphi.\\
\text{if } \lambda = -4, \, \alpha = -3 \\
X_4 = t^2 \partial_t + t \varphi \partial_\varphi
\end{array}$
\\ \hline
\end{tabular}
\end{adjustbox}
\end{table}

\textbf{Remark}. A polytropic gas corresponds to the function $G = -\varphi_s^\lambda$
\cite{bk:Chernyi_gas},
where $\lambda$ and the polytropic exponent $\gamma$ are related by the formula
$\lambda = -\gamma - 1$.
In particular, for the hyperbolic shallow water equations $\lambda = -3$
\cite{bk:SiriwatKaewmaneeMeleshko2016}.

\section{Conservation laws}

Noether's theorem allows one to find conservation
laws for the obtained equations using the Lagrangian (\ref{lagr})
and the found admitted symmetries \cite{bk:Ibragimov[1983]}.
The theorem states that if a Lagrangian $\mathcal{L}$
is invariant of the action of a symmetry $X$, i.e.,
\[
  X \mathcal{L} + \mathcal{L} \left( D_t \xi^t + D_s \xi^s \right) = D_t(V^t) + D_s(V^s),
\]
then the Euler-Lagrange equation (\ref{el}) possesses the following conservation law
\[
D_t(T^t) + D_s(T^s)
= D_t \left(
    \xi^t \mathcal{L} + \zeta \frac{\partial \mathcal{L}}{\partial \varphi_t} - V^t
\!\right)
+ D_s \left(
    \xi^s \mathcal{L} + \zeta \frac{\partial \mathcal{L}}{\partial \varphi_s} - V^s
\!\right) = 0,
\]
where $\zeta = \eta - \xi^t \varphi_t - \xi^s \varphi_s$.

If the vector $(V^t, V^s)$ is not a zero vector,
the conservation law is also called a divergent one.

Conservation laws in Lagrangian coordinates and their representations
in Eulerian coordinates are listed in Table~\ref{tab:cl_lagr} and Table~\ref{tab:cl_euler} respectively.
Note also that there are conservation laws in Lagrange coordinates
which have no representations in Eulerian coordinates.


\section{Conclusion}

Comprehensive analysis of equation (\ref{el}) is given in the present paper. Equation (\ref{el}) describes a motion of continuum (\ref{eq:main}) in Lagrangian coordinates. It contains two arbitrary functions $G(\varphi_s)$ and $H(\varphi$). Particular choices of these functions correspond to different models studied in continuum mechanics, such as,  isentropic flows of an ideal gas \cite{bk:Chernyi_gas}, different types of hyperbolic shallow water equations \cite{bk:SzatmariBihlo[2014],bk:Whitham[1974]}. One of the objectives of the present paper is to derive conservation laws of a continuum defined by equation (\ref{el}).

As equation (\ref{el}) is an Euler-Lagrange equation, for constructing conservation laws we used Noether's theorem. To apply Noether's theorem, first complete group analysis of equation (\ref{el}) is performed. Results of the group classification are given in Table \ref{tab:syms_arb} and Table \ref{tab:syms}. The group classification separates out the nonlinear function $G$ into the following forms: either the function $G$ is arbitrary or it has the exponential form $-e^{\mu\varphi_s}$ or the polynomial form $-\varphi_s^\lambda$, where $\mu$ and $\lambda$ are constant. All extensions of the kernels of admitted Lie algebras are found. These extensions occur for a particular cases of the function $H$.

Second using Noether's theorem, the conservation laws of equation (\ref{ele}) were obtained in Lagrangian. Their representations in Eulerian coordinates were also presented. It should be noted that some of the conservation laws have no their counterpart in Eulerian coordinates. All found conservation laws are listed in Table \ref{tab:cl_lagr} and Table \ref{tab:cl_euler}.

\section*{Acknowledgements}
The research  was supported by Russian Science Foundation Grant No 18-11-00238
`Hydrodynamics-type equations: symmetries, conservation laws, invariant difference schemes'.
E.I.K. also acknowledges Suranaree University of Technology for 
Full-time Master Researcher Fellowship (15/2561).
The authors thank V.A.Dorodnitsyn and E.Schulz for valuable discussions.




\begin{landscape}

\begin{longtable}[c]{c|c|ccc}
\caption{Conservation laws in Lagrangian coordinates}
\label{tab:cl_lagr}
\\
\hline
 $G$
 & $H$
 & $X$
 & conditions
 & conservation law $(T^t, T^s)$, where $g^{\prime\prime} \equiv G$, $h^\prime \equiv H$
\\
\hline
\hline
\endfirsthead

\multicolumn{5}{c}%
{\tablename\ \thetable\ -- \textit{Conservation laws in Lagrangian coordinates (continued)}} \\
\hline
\endhead

\endfoot

\endlastfoot

\multirow{12}{*}{$G(\varphi_s)$}
    & \multirow{2}{*}{$H(\varphi)$}
        & $\partial_t$ & - &
                $\left( \frac{1}{2} \varphi_t^2 - g - h, \,
                    \varphi_t g^\prime \right)$
            \\ 
        && $\partial_s$ & - &
                $\left( \varphi_t \varphi_s, \,
                        \varphi_s g^\prime - \frac{1}{2} \varphi_t^2 - g - h
                    \right)$
            \\ \cline{2-5}
    & \multirow{2}{*}{$0$}
        & $\partial_\varphi$ & - &
            $\left(\varphi_t, \, g^\prime \right)$
            \\ 
        && $t \partial_\varphi$ & - &
            $\left( t \varphi_t - \varphi, \, t g^\prime \right)$
            \\ \cline{2-5}
& \multirow{4}{*}{$\alpha \varphi$}
        & $\sin {\sqrt{|\alpha|} t} \partial_\varphi$
            & $\alpha < 0$ &
                $\left(
                        {\sqrt{|\alpha|} \varphi} \cos {\sqrt{|\alpha|} t}
                        - \varphi_t \sin {\sqrt{|\alpha|} t}
                    ,\,
                        -g^\prime \sin {\sqrt{|\alpha|} t}
                    \right)$
            \\ 
        && $\cos {\sqrt{|\alpha|} t} \partial_\varphi$
            & $\alpha < 0$ &
                $\left(
                    {\sqrt{|\alpha|} \varphi} \sin {\sqrt{|\alpha|} t}
                    + \varphi_t \cos {\sqrt{|\alpha|} t},
                \,
                    g^\prime \cos {\sqrt{|\alpha|} t}
                \right)$
              \\ 
        && $\sinh {\sqrt{\alpha} t} \partial_\varphi$
            & $\alpha > 0$ &
                $\left(
                        {\sqrt{\alpha} \varphi} \cosh {\sqrt{\alpha} t}
                        - \varphi_t \sinh {\sqrt{\alpha} t}, \,
                        -g^\prime \sinh {\sqrt{\alpha} t}
                    \right)$
            \\ 
        && $\cosh {\sqrt{\alpha} t} \partial_\varphi$
            & $\alpha > 0$  &
                    $\left(
                        {\sqrt{\alpha} \varphi} \sinh {\sqrt{\alpha} t}
                        - \varphi_t \cosh {\sqrt{\alpha} t}
                    , \,
                        -g^\prime \cosh {\sqrt{\alpha} t}
                    \right)$
            \\ \hline

    $\begin{array}{l}
        -e^{\mu \varphi_s}, \\
         \mu \neq 0
    \end{array}$
& \multirow{2}{*}{$0$}
     & $\begin{array}{l}
            3 \partial_t + s \partial_s
            + \left(\varphi - \frac{4}{\mu} s \right) \partial_\varphi
        \end{array}$
     & - &
        $\begin{array}{l}
        \Big(		
            \frac{3}{2}\,t{\varphi_{{t}}}^{2}
            - \left(
                \varphi -
                    \left( \varphi_{{s}}+4\,{\mu}^{-1} \right) s
                \right) \varphi_{{t}}
            +3\, t {{{{\rm e}^{\mu\,\varphi_{{s}}}}}{{\mu}^{-2}}}
		,\\
            -\frac{1}{2} \, s {\varphi_{{t}}}^{2}
            - \mu^{-2} {{\rm e}^{\mu\,\varphi_{{s}}}}
            \left[
                \left(
                    \mu \varphi_s + 3
                \right) s
                + \mu (3\, t \varphi_{t} - \varphi)
            \right]
             \Big)
        \end{array}$
        \\ \hline
\multirow{6}{*}{
$\begin{array}{l}
-(\varphi_s + c)^\lambda, \\
\lambda \neq 0, 1
\end{array}$
}
& \multirow{2}{*}{$0$}
    & $\begin{array}{l}
        (3 \lambda + 4) t \partial_t \\
            \quad + (\lambda + 4) s \partial_s
                + \lambda \varphi \partial_\varphi
        \end{array}$
        & - &
     $\begin{array}{l}
            \Big(		
                 \left(
                    \frac{3}{2}\,\lambda + 2
                 \right) t{\varphi_{{t}}}^{2}
                 + \left(
                    \left( \lambda +4 \right) \varphi_{{s}}s
                    - \lambda\,\varphi
                 \right) \varphi_{{t}}
                 \\
                 \qquad
                 + \left(
                        {\frac { 3\,\lambda + 4}
                        { \left( \lambda+1 \right)  \left( \lambda+2 \right) }} \, {\varphi_{{s}}}^{\lambda+2}
                 \right) t
			, \\
				\qquad \quad - (\lambda + 4) s \varphi_t^2 \left(
                        \frac{1}{\lambda + 2} \varphi_s^\lambda
                        + \frac{1}{2}
                    \right)
                \\
                \qquad \quad \quad
                - \frac{1}{\lambda + 1} \, t {\varphi_{{s}}}^{\lambda+1}
                    \left(
                        (3 \lambda + 4) \varphi_{{t}}
                        + \lambda \varphi
                    \right)
			\Big)
        \end{array}$
 \\ 
    && $ 2 t \partial_t + \varphi \partial_\varphi$ &
    $\lambda = -4$ &
     $\begin{array}{l}
            \Big(
				3 \,\varphi_{{t}} \left(
                    t \varphi_{{t}}- \varphi
                    \right)
                    + \, t {\varphi_{{s}}}^{-2} ,\,
                    {\varphi_{{s}}}^{-3} \left(
                    2 t \varphi_{{t}} - \varphi
                \right)
			\Big)
        \end{array}$
 \\ 
 && $\begin{array}{l}
        t^2 \partial_t + t \varphi \partial_\varphi
    \end{array}$ &
    $\lambda = -4$ &
     $\begin{array}{l}
            \Big(
                \left(
                    3 {\varphi_{{t}}}^{2} + {\varphi_{{s}}}^{-2}
                \right)
                {t}^{2}
                - 3 \,\varphi \left(
                    2 \,\varphi_{{t}}t - {\varphi}
			     \right)
            , \,
                2\, t {\varphi_{{s}}}^{-3} \left( t\varphi_{{t}} - \varphi \right)
			\Big)
        \end{array}$
 \\ \cline{2-5}
& \multirow{2}{*}{$\alpha \varphi$}
    & $2 s \partial_s - \varphi \partial_\varphi$
    &
            $\begin{array}{l}
            \lambda = -4/3, \\
            c = 0
            \end{array}$
     &
	   $\begin{array}{l}
            \Big(
				\left(
                    2 s \varphi_s + \alpha
                \right) \varphi_t
			, \,
            - \left(
                    9 \varphi_s^{2/3} + \varphi_t^2 + \alpha \varphi^2
                \right) s
                    - 3 \varphi_s^{-1/3} \left(
                        2 s \varphi_s + \varphi
                    \right)
			\Big)
        \end{array}$
    \\ 
        && $\begin{array}{l}
                \cos 2 \sqrt{|\alpha|} t \partial_t \\
                \quad - {\sqrt{|\alpha|} \varphi} \sin 2 \sqrt{|\alpha|} t \partial_\varphi
            \end{array}$
        & $\begin{array}{l}
            \lambda = -4, \\
            \alpha < 0
            \end{array}$
        &
        $\begin{array}{l}
            \Big(
                \cos 2\,\sqrt {|\alpha| }t \, \left(
                    3\,{\varphi_{{t}}}^{2}+3\,\alpha \,{\varphi}^{2} +{\varphi_{{s}}}^{-2}
                + 6\,\sqrt {|\alpha| }\varphi\, \varphi_{{t}} \sin 2\,\sqrt {|\alpha| }t
                \right)
            , \\ \qquad
                2 \varphi_s^{-3} \left(
                    \varphi_{{t}} \, \cos 2\,\sqrt {|\alpha| }t
                    + \sqrt {|\alpha| }\varphi \, \sin 2\,\sqrt {|\alpha| }t
                \right)
            \Big) \\
        \end{array}$
    \\ 
        && $\begin{array}{l}
                \sin 2 \sqrt{|\alpha|} t \partial_t \\
                \quad + {\sqrt{|\alpha|} \varphi} \cos 2 \sqrt{|\alpha|} t \partial_\varphi
            \end{array}$
        & $\begin{array}{l}
            \lambda = -4, \\
            \alpha < 0
            \end{array}$
        &
        $\begin{array}{l}
            \Big(
                \sin 2\,\sqrt {|\alpha| }t \, \left(
                    3\,{\varphi_{{t}}}^{2}+3\,\alpha \,{\varphi}^{2} +{\varphi_{{s}}}^{-2}
                - 6\,\sqrt {|\alpha| }\varphi\, \varphi_{{t}} \cos 2\,\sqrt {|\alpha| }t
                \right)
            , \\ \qquad
                2 \varphi_s^{-3} \left(
                    \varphi_{{t}} \, \sin 2\,\sqrt {|\alpha| }t
                    - \sqrt {|\alpha| }\varphi \, \cos 2\,\sqrt {|\alpha| }t
                \right)
            \Big) \\
        \end{array}$
    \\ 
        && $\begin{array}{l}
            \cosh 2 \sqrt{\alpha} t \partial_t \\
            \quad + {\sqrt{\alpha} \varphi} \sinh 2 \sqrt{\alpha} t \partial_\varphi
            \end{array}$
        & $\begin{array}{l}
            \lambda = -4, \\
            \alpha > 0
            \end{array}$
        &
        $\begin{array}{l}
            \Big(
                \cosh 2\,\sqrt {\alpha }t \, \left(
                    3\,{\varphi_{{t}}}^{2}+3\,\alpha \,{\varphi}^{2} +{\varphi_{{s}}}^{-2}
                - 6\,\sqrt {\alpha }\varphi\, \varphi_{{t}} \sinh 2\,\sqrt {\alpha }t
                \right)
            ,
            \\ \qquad
                2 \varphi_s^{-3} \left(
                    \varphi_{{t}} \, \cosh 2\,\sqrt {\alpha }t
                    - \sqrt {\alpha }\varphi \, \sinh 2\,\sqrt {\alpha }t
                \right)
            \Big) \\
        \end{array}$
    \\ 
        && $\begin{array}{l}
            \sinh 2 \sqrt{\alpha} t \partial_t \\
            \quad + {\sqrt{\alpha} \varphi} \cosh 2 \sqrt{\alpha} t \partial_\varphi
            \end{array}$
        & $\begin{array}{l}
            \lambda = -4, \\
            \alpha > 0
            \end{array}$
        &
        $\begin{array}{l}
            \Big(
                \sinh 2\,\sqrt {\alpha }t \, \left(
                    3\,{\varphi_{{t}}}^{2}+3\,\alpha \,{\varphi}^{2} +{\varphi_{{s}}}^{-2}
                - 6\,\sqrt {\alpha }\varphi\, \varphi_{{t}} \cosh 2\,\sqrt {\alpha }t
                \right)
            ,
            \\ \qquad
                2 \varphi_s^{-3} \left(
                    \varphi_{{t}} \, \sinh 2\,\sqrt {\alpha }t
                    - \sqrt {\alpha }\varphi \, \cosh 2\,\sqrt {\alpha }t
                \right)
            \Big) \\
        \end{array}$
    \\ \cline{2-5}
& \multirow{3}{*}{$\beta \varphi^\alpha$}
   & $\begin{array}{l}
        t \partial_t
        + \frac{\alpha + 3}{\alpha - 1} s \partial_s
        + \frac{2\varphi}{1 - \alpha}\partial_\varphi\end{array}$
        &
        $\begin{array}{l}
            \lambda = -\frac{8}{\alpha + 5}, \\
            \alpha \neq \pm1, 3, -5, \\
            c = 0
        \end{array}$
        &
    	$\begin{array}{l}
            \Bigg(
				\frac{t \varphi_t^2}{2}
                + \frac{((\alpha + 3) s \varphi_s + 2 \varphi ) \varphi_t}{\alpha - 1}
                + \frac{t}{\alpha + 1} \left(
                    \frac{(\alpha + 5)^2}{2 (\alpha - 3)} \, \varphi_s^{\frac{2(\alpha + 1)}{\alpha + 5}}
                    - \beta \varphi^{\alpha + 1}
                \right)
			, \\ \quad
                \frac{ t (\alpha + 5)}{(3 - \alpha)} \varphi_t \varphi_s^{\frac{\alpha - 3}{\alpha + 5}}
                - \frac{\alpha + 3}{2 (\alpha - 1)} s \varphi_t^2
                + \frac{2 (\alpha + 5)}{(3 - \alpha)(\alpha - 1)} \varphi \varphi_s^{\frac{\alpha - 3}{\alpha + 5}}
                \\
                 \qquad + \frac{(\alpha + 3) s}{2 (\alpha^2 - 1)} \left(
                    (\alpha + 5) \varphi_s^{\frac{ 2 (1 - \alpha)}{\alpha + 5}}
                    -  2 \beta \varphi^{\alpha + 1}
                \right)
			\Bigg)
        \end{array}$
    \\ 
        && $t^2 \partial_t + t \varphi \partial_\varphi$
        & $\begin{array}{l}
            \lambda = -4, \\
            \alpha = -3, \\
            c = 0
            \end{array}$
        &
        $\begin{array}{l}
        \Big(
            \left(
                3\,{\varphi_{{t}}}^{2}
                + 3 \beta {\varphi}^{-2}
                + {\varphi_{{s}}}^{-2}
            \right) {t}^{2}
            - 3 \varphi \left( 2 t \varphi_{{t}} - \varphi \right)
        ,
        \\
        \quad
            2\, t \varphi_s^{-3} \left( t \varphi_{{t}} - \varphi \right)
        \Big)
        \end{array}$
    \\ \hline
\end{longtable}




\centering
\begin{longtable}{c|c|ccc}
\caption{Conservation laws in Eulerian coordinates}
\label{tab:cl_euler}
\\
\hline
 $G$
 & $H$
 & $X$
 & conditions
 & conservation law $(T^t, T^x)$, where $g^{\prime\prime} \equiv G$, $h^\prime \equiv H$
\\
\hline
\hline
\endfirsthead

\multicolumn{5}{c}%
{\tablename\ \thetable\ -- \textit{Conservation laws in Eulerian coordinates (continued)}} \\
\hline
\endhead

\endfoot

\endlastfoot

\multirow{6}{*}{$G\left( \frac{1}{\rho} \right)$}
    & \multirow{2}{*}{$H(x)$}
        & $\partial_t$ & - &
                $\left(
                        \rho\, \left( {u}^{2}-2\,g - 2\,h \right)
                , \,
                    2 u \left(
                        g^\prime + \left( \frac{1}{2} u^2 - g - h \right) \rho
                    \right)
                \right)$
            \\ 
        && $\partial_s$ & - &
                $\left( u
                    , \,
                    \frac{1}{2} u^2 - g - h + \rho^{-1} g^\prime
                \right)$
            \\ \cline{2-5}
    & \multirow{2}{*}{$0$}
        & $\partial_\varphi$ & - &
            $\left(\rho u, \, \rho u^2 + g^\prime \right)$
            \\ 
        && $t \partial_\varphi$ & - &
            $\left( \rho\, \left( x - t u \right)
            , \,
                u\rho\, \left( x - tu \right)
                - t g^\prime
            \right)$
            \\ \cline{2-5}
& \multirow{2}{*}{$\alpha \varphi$}
        & $\sin {\sqrt{|\alpha|} t} \partial_\varphi$
            & $\alpha < 0$ &
                $\begin{array}{l}
                    \Big(
                        \rho (
                            \sqrt{|\alpha|} x \cos {\sqrt{|\alpha|} t}
                            - u \sin {\sqrt{|\alpha|} t}
                        )
                    ,
                    \\
                    \qquad
                        x \rho u \sqrt{|\alpha|} \cos {\sqrt{|\alpha|} t}
                        - (u^2 \rho + g^\prime) \sin{\sqrt{|\alpha|} t}
                    \Big)
                    \end{array}$
            \\ 
        && $\cos {\sqrt{|\alpha|} t} \partial_\varphi$
            & $\alpha < 0$ &
                $\begin{array}{l}
                    \Big(
                        \rho (
                            \sqrt{|\alpha|} x \sin {\sqrt{|\alpha|} t}
                            + u \cos {\sqrt{|\alpha|} t}
                        )
                    ,
                    \\ \qquad
                        x \rho u \sqrt{|\alpha|} \sin {\sqrt{|\alpha|} t}
                        + (u^2 \rho + g^\prime) \cos{\sqrt{|\alpha|} t}
                    \Big)
                    \end{array}$
              \\ 
        && $\sinh {\sqrt{\alpha} t} \partial_\varphi$
            & $\alpha > 0$ &
                $\begin{array}{l}
                    \Big(
                        \rho (
                            \sqrt{\alpha} x \cosh {\sqrt{\alpha} t}
                            - u \sinh {\sqrt{\alpha} t}
                        )
                    ,
                    \\
                    \qquad
                        x \rho u \sqrt{\alpha} \cosh {\sqrt{\alpha} t}
                        - (u^2 \rho + g^\prime) \sinh {\sqrt{\alpha} t}
                    \Big)
                    \end{array}$
            \\ 
        && $\cosh {\sqrt{\alpha} t} \partial_\varphi$
            & $\alpha > 0$  &
                    $\begin{array}{l}
                    \Big(
                        \rho (
                            \sqrt{\alpha} x \sinh {\sqrt{\alpha} t}
                            - u \sinh {\sqrt{\alpha} t}
                        )
                    ,
                    \\
                    \qquad
                        x \rho u \sqrt{\alpha} \sinh {\sqrt{\alpha} t}
                        - (u^2 \rho + g^\prime) \cosh {\sqrt{\alpha} t}
                    \Big)
                    \end{array}$
            \\ \hline
\multirow{6}{*}{
$\begin{array}{l}
-(\varphi_s + c)^\lambda, \\
\lambda \neq 0, 1
\end{array}$
}
& \multirow{2}{*}{$0$}
    & $ 2 t \partial_t + \varphi \partial_\varphi$ &
    $\lambda = -4$ &
     $\begin{array}{l}
            \Big(
				\rho\, \left( \left( {u}^{2} + \frac{1}{3} \,{\rho}^{2} \right) t - \,ux \right)
			, \\ \qquad
				\rho\, \big(
                    u \left( {\rho}^{2}+{u}^{2}\right) t
                    - \left( {u}^{2}+\frac{1}{3} \,{\rho}^{2} \right) x
                \big)
			\Big)
        \end{array}$
 \\ 
 && $ t^2 \partial_t + t \varphi \partial_\varphi
    $
    &
    $\lambda = -4$ &
     $\begin{array}{l}
            \Big(
				\rho\,(
                    (\frac{1}{3}\,\rho\,^2 + u\,^2)\,t^2
                    - x\,(2\,t\,u - x)
                )
            \, \\ \qquad
                \rho \big(
                    (\rho^2 + u^2) u t^2
                    - 2 x (\frac{1}{3} \rho^2+u^2) t
                    + u x^2
                \big)
			\Big)
        \end{array}$
 \\ \cline{2-5}
& \multirow{2}{*}{$\alpha \varphi$}
        & $\begin{array}{l}
                \cos 2 \sqrt{|\alpha|} t \partial_t \\
                \quad - {\sqrt{|\alpha|} \varphi} \sin 2 \sqrt{|\alpha|} t \partial_\varphi
            \end{array}$
        & $\begin{array}{l}
            \lambda = -4, \\
            \alpha < 0
            \end{array}$
        &
        $\begin{array}{l}
            \Big(
                \rho \left(
                    x u \sqrt{|\alpha|} \sin {2 \sqrt{|\alpha| t}}
                    + \frac{1}{2} \left( u^2 -  |\alpha| x^2 + \frac{1}{3} \rho^2 \right) \cos {2 \sqrt{|\alpha|} t}
                \right)
            ,
            \\ \qquad
                \rho \left(
                    \sqrt{|\alpha|} x \left( u^2+ \frac{1}{3} \rho^2 \right) \sin{2 \sqrt{|\alpha|} t}
                    + \frac{1}{2} \, u \, (u^2 - |\alpha| x^2 + \rho^2) \cos {2 \sqrt{|\alpha|} t}
                \right)
            \Big) \\
        \end{array}$
    \\ 
        && $\begin{array}{l}
                \sin 2 \sqrt{|\alpha|} t \partial_t \\
                \quad + {\sqrt{|\alpha|} \varphi} \cos 2 \sqrt{|\alpha|} t \partial_\varphi
            \end{array}$
        & $\begin{array}{l}
            \lambda = -4, \\
            \alpha < 0
            \end{array}$
        &
        $\begin{array}{l}
            \Big(
                \rho \left(
                    x u \sqrt{|\alpha|} \cos {2 \sqrt{|\alpha| t}}
                    - \frac{1}{2} \left( u^2 -  |\alpha| x^2 + \frac{1}{3} \rho^2 \right) \sin {2 \sqrt{|\alpha|} t}
                \right)
            ,
            \\ \qquad
                \rho \left(
                    \sqrt{|\alpha|} x \left( u^2+ \frac{1}{3} \rho^2 \right) \cos {2 \sqrt{|\alpha|} t}
                    - \frac{1}{2} \, u \, (u^2 - |\alpha| x^2 + \rho^2) \sin {2 \sqrt{|\alpha|} t}
                \right)
            \Big) \\
        \end{array}$
    \\ 
        && $\begin{array}{l}
            \cosh 2 \sqrt{\alpha} t \partial_t \\
            \quad + {\sqrt{\alpha} \varphi} \sinh 2 \sqrt{\alpha} t \partial_\varphi
            \end{array}$
        & $\begin{array}{l}
            \lambda = -4, \\
            \alpha > 0
            \end{array}$
        &
        $\begin{array}{l}
            \Big(
                \rho \left(
                    x u \sqrt{\alpha} \sinh {2 \sqrt{\alpha t}}
                    - \frac{1}{2} \left( u^2 +  \alpha x^2 + \frac{1}{3} \rho^2 \right) \cosh {2 \sqrt{\alpha} t}
                \right)
            ,
            \\ \qquad
                \rho \left(
                    \sqrt{\alpha} x \left( u^2+ \frac{1}{3} \rho^2 \right) \sinh {2 \sqrt{\alpha} t}
                    - \frac{1}{2} \, u \, (u^2 + \alpha x^2 + \rho^2) \cosh {2 \sqrt{\alpha} t}
                \right)
            \Big) \\
        \end{array}$
    \\ 
        && $\begin{array}{l}
            \sinh 2 \sqrt{\alpha} t \partial_t \\
            \quad + {\sqrt{\alpha} \varphi} \cosh 2 \sqrt{\alpha} t \partial_\varphi
            \end{array}$
        & $\begin{array}{l}
            \lambda = -4, \\
            \alpha > 0
            \end{array}$
        &
        $\begin{array}{l}
            \Big(
                \rho \left(
                    x u \sqrt{\alpha} \cosh {2 \sqrt{\alpha t}}
                    - \frac{1}{2} \left( u^2 +  \alpha x^2 + \frac{1}{3} \rho^2 \right) \sinh {2 \sqrt{\alpha} t}
                \right)
            , \\ \qquad
                \rho \left(
                    \sqrt{\alpha} x \left( u^2+ \frac{1}{3} \rho^2 \right) \cosh {2 \sqrt{\alpha} t}
                    - \frac{1}{2} \, u \, (u^2 + \alpha x^2 + \rho^2) \sinh {2 \sqrt{\alpha} t}
                \right)
            \Big) \\
        \end{array}$
    \\ \cline{2-5}
& \multirow{1}{*}{$\beta \varphi^\alpha$}
        & $t^2 \partial_t + t \varphi \partial_\varphi$
        & $\begin{array}{l}
            \lambda = -4, \\
            \alpha = -3, \\
            c = 0
            \end{array}$
        &
        $\begin{array}{l}
        \Big(
            \rho \left( \left(\frac{1}{3} \rho^2+u^2
                + \beta x^{-2} \right) t^2
                - x (2 t u-x) \right)
        ,
        \\
        \qquad
            \rho \left(
                (\rho^2 + u^2+ \beta x^{-2}) t^2 u
                - 2 tx \left( \frac{1}{3} \rho^2+u^2 \right)
                + u x^2
            \right)
        \Big)
        \end{array}$
    \\ \hline
\end{longtable}

\end{landscape}

\end{document}